\documentclass[reprint,amsmath,amssymb,aps,prb,groupedaddress,nofootinbib,twocolumn,superscriptaddress]{revtex4-2}
\usepackage{graphicx}
\usepackage{amsthm,amssymb,amsmath,braket,mathdots}
\usepackage{bm}
\usepackage[pagebackref=false,pdfnewwindow=true]{hyperref} 
\usepackage{epstopdf,psfrag}
\usepackage{relsize,amsbsy}
\usepackage[export]{adjustbox}

\usepackage{graphicx,xcolor} 
\usepackage{tabularx, booktabs} 

\usepackage{units} 
\usepackage{dsfont} 

\newcommand{\ii}{{\mathrm{i}}} 
\newcommand{\e}{{\mathrm{e}}} 
\renewcommand{\v}[1]{\bm{#1}} 

\newcommand{\mac}{\mathcal}

\renewcommand{\eqref}[1]{Eq.~(\ref{#1})}

\usepackage{tikz}
\usetikzlibrary{arrows.meta}
\newcommand{\tstar}[5]{
\pgfmathsetmacro{\starangle}{360/#3}
\draw[#5] (#4:#1)
\foreach \x in {1,...,#3}
{--(#4+\x*\starangle-\starangle/2:#2) -- (#4+\x*\starangle:#1)}
-- cycle;}

\definecolor{bananayellow}{rgb}{1.0, 0.88, 0.21}
\definecolor{straw}{rgb}{0.32, 0.28, 0.1}

\usepackage[caption=false]{subfig}
\usepackage{changes}
\definechangesauthor[name=Valentin, color=blue]{VL} 
\definechangesauthor[name=Johannes, color=red]{JK}

\begin{document}

\title{Interband scattering- and nematicity-induced quantum oscillation frequency in FeSe}
\author{Valentin Leeb}
\affiliation{Technical University of Munich, TUM School of Natural Sciences, Physics Department, TQM, 85748 Garching, Germany}
\affiliation{Munich Center for Quantum Science and Technology (MCQST), Schellingstr. 4, 80799 M{\"u}nchen, Germany}
\author{Johannes Knolle}
\affiliation{Technical University of Munich, TUM School of Natural Sciences, Physics Department, TQM, 85748 Garching, Germany}
\affiliation{Munich Center for Quantum Science and Technology (MCQST), Schellingstr. 4, 80799 M{\"u}nchen, Germany}
 	\affiliation{\small Blackett Laboratory, Imperial College London, London SW7 2AZ, United Kingdom}
\date{\today}

\begin{abstract}
Understanding the nematic phase observed in the iron-chalcogenide materials is crucial for describing their superconducting pairing. Experiments on FeSe$_{1-x}$S$_x$ showed that one of the slow Shubnikov--de Haas quantum oscillation frequencies disappears when tuning the material out of the nematic phase via chemical substitution or pressure, which  has been interpreted as a Lifshitz transition [Coldea {\it et al}., npj Quant Mater 4, 2 (2019), Reiss {\it et al}., Nat. Phys. 16, 89–94 (2020)]. Here, we present a generic, alternative scenario for a nematicity-induced sharp quantum oscillation frequency which disappears in the tetragonal phase and is not connected to an underlying Fermi surface pocket. We show that different microscopic interband scattering mechanisms -- for example, orbital-selective scattering -- in conjunction with nematic order can give rise to this quantum oscillation frequency beyond the standard Onsager relation. We discuss implications for iron-chalcogenides and the interpretation of quantum oscillations in other correlated materials.  
\end{abstract}
	
\maketitle

{\it Introduction.--} The availability of experimental methods, which are able to correctly identify the low energy electronic structure of quantum materials, is critical for understanding their emergent phenomena like superconductivity, various density waves or nematic orders. For example, angle-resolved photoemission spectroscopy (ARPES) on the cuprate materials confirmed that a single band Hubbard-like description is a reasonable starting point for modelling their low energy structure~\cite{damascelli2003angle}, but iron-based superconductors require a  multi-band, multi-orbital description~\cite{lu2008electronic,richard2015arpes,yi2017role}. Beyond ARPES, quantum oscillation (QO) measurements are an exceptionally sensitive tool for measuring Fermi surface (FS) geometries as well as interaction effects via extracting the effective masses from the temperature dependence~\cite{Shoenberg}. For example, QO studies famously confirmed the presence of a  closed FS pocket in underdoped cuprates in a field~\cite{doiron2007quantum,sebastian2012towards} or observed the emergence of small pockets in the spin density wave parent phase of iron-based superconducting compounds~\cite{sebastian2008quantum,terashima2011complete,coldea2013iron}.

The interpretation of QOs, as measured in transport or thermodynamic observables, is based on the famous Onsager relation, which ascribes each QO frequency to a semi-classical FS orbit~\cite{Onsager1952,Shoenberg}. In the past years, this canonical description has been challenged by the observation of {\it anomalous} QOs in correlated insulators~\cite{tan2015unconventional,czajka2021oscillations} which motivated a number of works revisiting the basic theory of QOs~\cite{knolle2015quantum,zhang2016quantum,knolle2017excitons,sodemann2018quantum,erten2016kondo,Chowdhury2018,shen2018quantum,lee2021quantum,leeb2021anomalous,Allocca2021, Allocca2022, Allocca2023,leeb2023quantum}. Very recently, forbidden QO frequencies have been reported in the multi-fold semi-metal CoSi~\cite{Huber2023}, which generalize so-called magneto-intersubband oscillations known in coupled 2D electron gases~\cite{Polyanovsky1988,Raikh1994,Averkiev2001} to generic bulk metals~\cite{Leeb_DiffFreq}. In Ref.~\cite{Huber2023} it was proposed that QO of the quasiparticle lifetime in systems with multiple allowed FS orbits can lead to new combination frequencies without a corresponding semi-classical FS trajectory.

Here, we propose a new explanation for the QO spectra measured in the iron-chalcogenide superconductor FeSe$_{1-x}$S$_x$ which leads to an alternative identification of its low energy electronic structure with direct implications for the superconducting pairing. Iron-chalcogenides are unique among the iron-based superconductors as they show an orthorombic distortion without stripe magnetism, i.e. pristine FeSe is already in a nematic phase~\cite{watson2015dichotomy,kasahara2014field,Terashima2014_Anomalous}. Recently it was reported that one of the observed slow QO frequencies (labeled as $\lambda$ in the experimental data) vanishes when tuning out of the nematic into the tetragonal phase, via pressure in FeSe$_{0.89}$S$_{0.11}$~\cite{Reiss2020} or via isoelectronic substitution in FeSe$_{1-x}$S$_x$ \cite{Coldea2019}. Following Onsager's standard theory it has been interpreted as a Lifshitz transition, i.e. a FS pocket present in the nematic phase which disappears at the nematic quantum critical point~\cite{Reiss2020}. As an alternative scenario, we show here that an additional slow QO frequency without an underlying FS orbit can naturally appear in an electronic nematic phase. 

\begin{figure}
	\centering
	\includegraphics[width=\columnwidth]{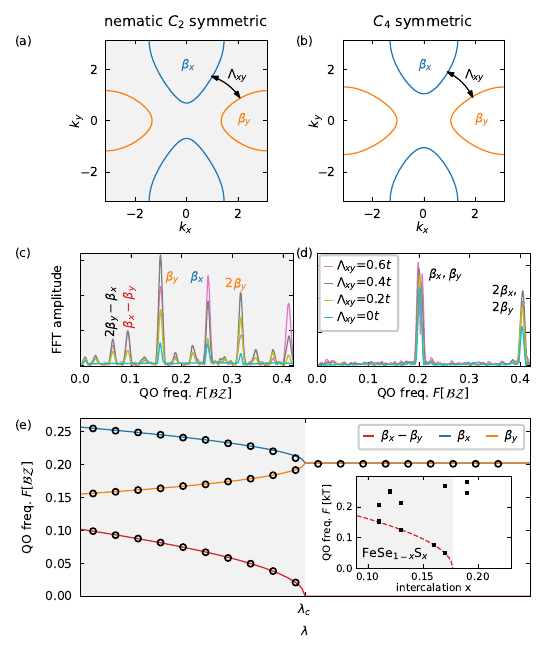}
	\caption{(b) FS of a minimal model including only 2 electron pockets $\beta_x,\beta_y$ with different orbital character. (a) In the nematic phase the $\beta_x$ pocket spontaneously grows whereas  $\beta_y$ shrinks, see (c) and (d) for representative numerical SdH QO spectra for the different phases. (e) In the nematic phase (gray background) the degenerate frequency of the C$_4$ symmetric phase splits up into two frequencies (blue, orange), each associated with one FS. When taking interband coupling from impurities ($\Lambda_{xy}$), see (a), (b) into account a third frequency (red) which is exactly the difference of the basis frequencies $\beta_x-\beta_y$ appears in the nematic phase. We fixed $t_1/2 = t_2 = t$, $\mu=-3t$. The inset of panel (e) shows the experimentally detected peak frequencies in FeSe$_{1-x}$S$_x$~\cite{Coldea2019}. The red dashed line ($\propto \sqrt{\lambda_c-\lambda}$) is a guide to the eye highlighting the emergent frequency in the nematic phase, identified as $\lambda$ in Ref.~\cite{Coldea2019,Reiss2020}.}
	\label{fig:1}
\end{figure}

Our scenario requires the following features of iron-chalcogenides~\cite{coldea2018key,baek2015orbital,shimojima2014lifting,occhialini2023spontaneous, Coldea2021_electronic}:
(i) The FS consists of several pockets, in particular two electron pockets (labeled here as $\beta_x$ and $\beta_y$) around the $Y$ and $X$ point of the Brillouin zone (BZ), see Fig.~\ref{fig:1} panel~(b). $\beta_x$ ($\beta_y$) has almost pure $d_{xz}$($d_{yz}$) orbital character with some $d_{xy}$ content. They are related to each other via a C$_4$ rotation in the tetragonal phase. (ii) When tuning  into nematic phase with broken rotational symmetry (reduced to C$_2$) one of the pockets spontaneously increases in size, whereas the other one shrinks, see panel~(a). In the QO spectrum, this is visible by the split up of one formerly degenerate QO frequency into 2 frequencies. (iii) A strong inter-pocket scattering between the $\beta_x$ and $\beta_y$ pocket exists~\cite{ortenzi2009fermi,breitkreiz2013transport,watson2015dichotomy,koshelev2016magnetotransport}. It can be caused either by orbital selective impurity scattering over the $d_{xy}$-channel, low-momentum scattering, collective fluctuations or, most likely, a combination of all. As a result, we will show that a new slow QO frequency, set by the difference of the $\beta_x$ and $\beta_y$ frequencies, emerges. 

We argue that our theory can not only explain the slow SdH QO frequency observed in iron-chalcogenides, but also discuss that it provides further support for the robustness of $s_{\pm}$ superconducting pairing. 

We note that we do not aim towards a full quantitative description of the complicated QO spectrum of FeSe but rather focus on presenting a new theory for the additional slow QO frequency appearing in the nematic phase, thus, concentrating on  model descriptions with the minimal ingredients of the electronic structure (e.g. neglecting aspects of three-dimensionality). 

The paper is organized as follows: We first introduce a basic two-band model which captures the minimal features of an electronic nematic phase transition. We then show that inter-pocket scattering leads to a new QO frequency in a full lattice calculation of the SdH effect, including the orbital magnetic field via Peierls substitution. Next, we discuss a more microscopic multi-orbital description of iron-chalcogenides and identify different scattering mechanisms leading to strong inter-electron pocket coupling. Again, we confirm the emergence of a slow nematicity-induced frequency in a full lattice calculation. We close with a summary and outlook. 


{\it Minimal two-pocket model.--} First, we consider a minimal model with two electron pockets and the Hamiltonian
\begin{equation}
H_0 = \sum_{\v{k}} (\epsilon_{x,\v{k}}-\delta \mu) d^\dagger_{x,\v{k}} d_{x,\v{k}} + (\epsilon_{y, \v{k}} +\delta\mu)d^\dagger_{y,\v{k}} d_{y,\v{k}}
\end{equation}
with the dispersion
$\epsilon_{x,\v{k}} = -2t_1 \cos k_x + 2t_2 \cos k_y$ and 
$\epsilon_{y,\v{k}} = 2t_2 \cos k_x - 2t_1 \cos k_y$. 
It consists of a $\beta_x$-FS pocket around the $Y$-point and a $\beta_y$-Fermi pocket around the $X$-point, see Fig.~\ref{fig:1}~(b). For $\delta \mu = 0$ the Hamiltonian is invariant under the $C_4$ rotation $(k_x,k_y) \!\! \rightarrow \!\! (k_y,-k_x), (d_x,d_y) \!\! \rightarrow \!\! (d_y,-d_x)$. 
Additional  density-density interactions $\sum_{\v{r},\alpha,\beta} d_{\alpha,\v{r}}^\dag d_{\alpha,\v{r}} d_{\beta,\v{r}}^\dag d_{\beta,\v{r}}$ can induce a nematic transition with a finite orbital asymmetry $\delta \mu \neq 0$ breaking the $C_4$ rotation symmetry. Mean-field calculations confirm  that $\delta \mu$ becomes non-zero for interactions above a critical threshold~\cite{Daghofer2010_three}.  Thus, $\delta \mu$ serves as an order parameter for a nematic phase transition, which is manifest in the band structure by the spontaneous growth/shrinking of the two inequivalent pockets, see Fig.~\ref{fig:1}~(a). We note that additional FS pockets are present in FeSe and change properties of the nematic phase quantitatively but are not relevant for our purpose.  

In practice an external parameter $\lambda$ tunes the effective interaction strength, e.g. via a change of applied pressure \cite{Reiss2020} or chemical substitution~\cite{Coldea2019}. Again, the precise relation between $\delta \mu(\lambda)$ and $\lambda$ depends on microscopic details but we assume in the following the generic form of a second order phase transition $\delta \mu \propto (\lambda_c-\lambda)^\alpha \theta(\lambda_c-\lambda)$ and fix, for simplicity, the exponent to be of the standard mean-field behaviour $\alpha=1/2$. 





Following our recent works~\cite{Huber2023,Leeb_DiffFreq}, we introduce a scattering contribution between the two electron pockets via impurities  
\begin{equation}
H_\text{imp} = \sum_{\v{r}} \Lambda_{xy,\v{r}} d^\dag_{x,\v{r}} d_{y,\v{r}} + h.c.
\end{equation}
where $\Lambda_{xy,\v{r}}$ are drawn randomly, independently and uniformly in space from the interval $[-\Lambda_{xy}/2,\Lambda_{xy}/2]$. On average the system retains its translation and rotation symmetry. For simplicity we set the intraorbital part of the impurities, i.e. $\Lambda_{xx}$ and $\Lambda_{yy}$ to zero, as they will only suppress the amplitude of all QO frequencies~\cite{Leeb_DiffFreq}.

We include a magnetic field by standard Peierls substitution, effectively inserting a flux $\Phi$ in each plaquette of the square lattice. We have implemented the hopping Hamiltonian with magnetic field and impurities for system sizes up to $300\times 300$ lattice sites. We determined the conductance through the Landau--B\"uttiker algorithm using the python package kwant~\cite{kwant} and observed SdH oscillations of the conductance as function of $1/\Phi$. We then analyzed the Fourier transformation in $2\pi/\Phi$ with standard QO techniques, which include subtraction of a polynomial background, zero padding and windowing, see SM. Representative Fourier spectra for the tetragonal (C$_4$ symmetric) and nematic phase are shown in Fig.~\ref{fig:1}~(c) and (d), where the frequencies are shown in units of the area of the BZ.

The Fourier spectrum of the SdH oscillations features, as expected from Onsager's relation, peaks at frequencies $F_\beta = S_\beta /2\pi e$, which correspond to the area of the respective FSs $S_\beta$ and higher harmonics thereof. As our main finding, the spectrum has clear peaks at combination frequencies in the nematic phase, most dominantly $\beta_x-\beta_y$. Crucially, this frequency does not have an underlying FS or semiclassical orbit of any kind but is a consequence of QO of the quasi-particle lifetime. We note that this is in accordance with our recent analytical work~\cite{Leeb_DiffFreq}, which we confirm here for the first time in a numerical lattice calculation. 

In Fig.~\ref{fig:1}~(e), we plot the frequencies of the 3 strongest signals for weak inter-orbit scattering as a function of the external parameter $\lambda$ tuning through the nematic transition. When increasing the nematic order, the main frequency peak splits into two, and the additional low-frequency $\beta_x-\beta_y$ oscillation emerges similar to the experimental data, see inset.

\begin{figure}
    \centering
    \includegraphics[width=\columnwidth]{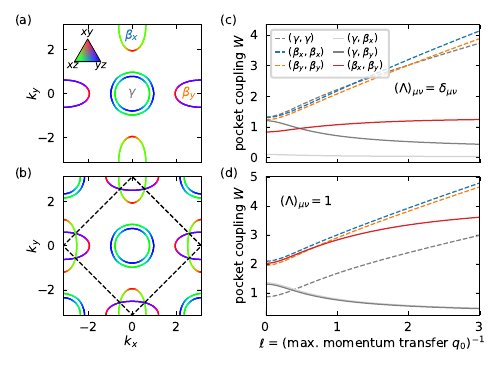}
    \caption{Panel (a): Typical FS of the three-orbital model in the nematic phase. Colors indicate the orbital character. Panel (b): Buckling enlarges the unit cell which leads to a backfolded FS in the reduced Brillouin zone (black dashed). Panel (c): FS integrated inter- and intrapocket scattering strength for $(\Lambda)_{\mu\nu} = \delta_{\mu\nu}$ showing that the coupling $W_{\beta_x,\beta_y}$ is dominant. It increases for small momentum scattering, i.e. $1/q_0 \rightarrow \infty$ . Any other type of interorbit scattering enlarges the coupling of $\beta_x$ and $\beta_y$ even further, see panel (d) where $(\Lambda)_{\mu\nu} = 1$.}
    \label{fig:2}
\end{figure}

{\it Multi-orbital model.--} After studying a minimal two-band model, we next want to understand the possible origin of a strong inter-pocket scattering. Therefore, we need to take the multi-orbital character of iron-chalcogenides into account. In order to keep the numerical lattice calculations tractable we focus on the following key features, see Fig.~\ref{fig:2}~(a): (i) Two electron-like elliptical pockets $\beta_x$ and $\beta_y$ around the $Y$ and $X$ points which have mainly $d_{xz}$ and $d_{yz}$ orbital character but in addition also an admixture of $d_{xy}$ orbitals; (ii) One (or depending on the precise model and parameter regime also two) hole-like circular pockets $\gamma$ around the $\Gamma$ point which have mixed $d_{xz}$ and $d_{yz}$ orbital character; (iii) Only the electron pockets $\beta_x$ and $\beta_y$ have additional $d_{xy}$ orbital character. 

All features (i)-(iii) are captured by a three orbital model~\cite{Daghofer2010_three} with $d_{xz}, d_{yz},$ and $d_{xy}$ orbitals (denoted by $xz,yz,xy$). Introducing $\v{\Psi}_{\v{k}} = (d_{\v{k}, xz},d_{\v{k}, yz},d_{\v{k}, xy})$, the Hamiltonian reads 
\begin{align}
H_0 &= \sum_{\v{k}} \v{\Psi}_{\v{k}}^\dagger (T(\v{k})-\mu) \v{\Psi}_{\v{k}} + \delta \mu
\begin{pmatrix}
d_{\v{k}, xz} \\ d_{\v{k}, yz}
\end{pmatrix}^\dag
\sigma^z
\begin{pmatrix}
d_{\v{k}, xz} \\ d_{\v{k}, yz}
\end{pmatrix}
\label{eq:H_3orbital}
\end{align}
where $T(\v{k})$ is a $3\times 3$ matrix which depends on the electronic hopping strengths between the orbitals. The real-space form of the Hamiltonian, $T(\v{k})$ and the parameters are given in the SM.

In the tetragonal phase, with $\delta \mu=0$, the Hamiltonian is again invariant under the $C_4$ rotation $(k_x,k_y) \rightarrow (k_y,-k_x), (d_x,d_y) \rightarrow (d_y,-d_x)$. Similar to the toy model from above, a nematic phase is characterized by a finite $\delta \mu$ where the rotation symmetry is reduced to a $\mathbb{Z}_2$ reflection symmetry / $C_2$ rotation symmetry.

The parameter $\delta \mu$ is again an effective, emergent parameter but now we can relate its microscopic origin to orbital ordering. For example the interorbital density interaction between $xz$ and $yz$ orbitals
\begin{equation}
H_\text{int} = U \sum_{r} d_{\v{r},xz}^\dag d_{\v{r},xz} d_{\v{r},yz}^\dag d_{\v{r},yz}.
\end{equation}
 can be decoupled in mean-field to obtain a self-consistent order parameter for the nematic (now orbital ordering) transition leading to $\delta \mu = U \left(\langle d_{\v{r},xz}^\dag d_{\v{r},xz} \rangle - \langle d_{\v{r},yz}^\dag d_{\v{r},yz} \rangle\right)/2$. A typical FS within the nematic phase is shown in Fig.~\ref{fig:2}~(a).

We note that this role of orbital ordering, or an imbalance of the orbital occupation, in the nematic phase has been confirmed in a number of experiments~\cite{baek2015orbital,shimojima2014lifting} most recently via X-ray linear dichroism~\cite{occhialini2023spontaneous}. While our minimal three-orbital model captures the key features, the precise asymmetry of the $\gamma$ hole pocket(s) in the nematic phase of FeSe is more complicated however its shape does not affect our new findings.


{\it Impurities and orbital selective scattering.--} As confirmed in our two-band model numerically and expected from analytical calculations~\cite{Leeb_DiffFreq}, a nematicity induced difference frequency requires a sizeable coupling of the pockets $\beta_x$ and $\beta_y$. The absence of other frequency combinations points towards a negligible coupling of $\beta_i$ and $\gamma$. We next investigate the origin of this coupling in terms of the $d$-orbital dependent scattering. Therefore, we consider impurities in the orbital basis
\begin{equation}
H_\text{imp} = \sum_{\v{r}} 
 \sum_{\v{r}_i} V(\v{r}-\v{r}_i) \v{\Psi}^\dag_{\v{r}} \Lambda_{\v{r}_i} \v{\Psi}_{\v{r}} 
\end{equation}
with the scattering vertex $\Lambda_{\v{r}_i}$ a random hermitian matrix with mean 0 and variance $\Lambda^2$. Note, impurities respect the $\pi/2$-rotation symmetry only on average. Similarly, impurities located at $\v{r}_i$ are distributed randomly and uniformly such that the systems remains on average
translationally invariant. 
We model the interaction of electrons with impurities by a screened Coulomb interaction $V_\ell$ of Yukawa type with screening length $\ell$ \cite{Flensberg}.

We quantify the coupling $W_{\alpha,\alpha'}$ of FS orbits $\alpha$ and $\alpha'$ by integrating the scattering amplitudes of all possible processes between them
\begin{align}
W_{\alpha,\alpha'} &= \oint_{\v{k} \in \alpha} \oint_{\v{k}' \in \alpha'} \Big|
\begin{tikzpicture}[baseline={(0,-0.07)}]
\draw [-{Stealth[length=3mm, width=2mm]}] (0.4,0) -- (0.52,0) node[xshift=-4pt,anchor=north]{\footnotesize $\v{k}$};
\draw [-{Stealth[length=3mm, width=2mm]}] (1.28,0) -- (1.4,0) node[anchor=north]{\footnotesize $\v{k}'$};
\draw (0,0) -- (1.7,0);
\fill (0.85,0) circle (0.07);
\draw[dashed] (0.85,0) -- +(0,0.7) node[yshift=-7pt,anchor=west]{\footnotesize $\v{k}'-\v{k}$};
\tikzset{shift={(0.85,0.7)}}
\tstar{0.03}{0.1}{5}{20}{fill=black}
\end{tikzpicture} \Big| \\
&=  \oint_{\v{k} \in \alpha} \oint_{\v{k}' \in \alpha'} \big| \tilde{V}_\ell(\v{k}'-\v{k}) \mac{U}(\v{k}')^\dag\Lambda \mac{U}(\v{k}) \big|.
\end{align}
Here,  $\mac{U}(\v{k})$ is the transformation which diagonalizes $H_{0}$ for a each momentum.
The Fourier transform of the screened Coulomb interaction $\tilde{V}_\ell = \mathcal{N}_\ell/(\v{k}^2 + 1/\ell^2)$ allows only scattering up to a maximal momentum  $q_0 = 1/\ell$ ($\mathcal{N}_\ell$ is a normalization constant).

Iron-chalcogenides have a 2 site unit cell~\cite{coldea2018key}, which leads to a folding of the $T(\v{k}+(\pi,\pi))$ bands onto the $T(\v{k})$ bands. The FS in the reduced Brillouin zone is shown in Fig.~\ref{fig:2}~(b), where now the pockets $\beta_x$ and $\beta_y$ lay on top of each other. This admits a large scattering between the $\beta_x$ and $\beta_y$ pockets because the screened Coulomb interaction favors low-momentum scattering. In Fig.~\ref{fig:2}~(c) and (d) we show quantitatively that for diagonal or uniform scattering vertices $\Lambda$ in the orbital components, the coupling $W_{\beta_x, \beta_y}$ is the biggest inter-pocket coupling for a sizeable screening length $\ell \gtrsim 0.5$ and of the same size as the intra-orbit couplings.

There are several additional mechanisms which increase $W_{\beta_x, \beta_y}$ even further. Crucially, orbital-selective scattering, i.e. a dominating $\Lambda_{xy,xy}$ component of the vertex, leads to a large coupling of exclusively $\beta_x$ and $\beta_y$ pockets. Additionally, any off-diagonal element of $\Lambda$, i.e. $xz/yz$ to $xy$ and $xz$ to $yz$ scattering, strongly enhances the inter-pocket coupling $W_{\beta_x, \beta_y}$. Overall, there is generically a sizeable coupling between the electron pockets.

An exclusive coupling of the electron pockets $\beta_x,\beta_y$ can be modelled by orbital selective scattering over the $\Lambda_{xy,xy}$ channel. The analysis above suggests that this coupling is indeed dominating. For our numerical simulation of the SdH effect we, therefore, focus on short-ranged impurities $V(\v{r}) \propto \delta(\v{r})$ with an orbital selective scattering vertex $(\Lambda)_{ij} = \delta_{i3} \delta_{j3} \Lambda_{xy}$ with only the $xy$ component $\Lambda_{xy,xy}$ being non-zero.
We note that experiments indeed suggest that the $xy$-orbital part of the FS is heavy, leading to a large dominating density of $d_{xy}$-states for scattering~\cite{coldea2018key}.

\begin{figure}
    \centering
    \includegraphics[width=\columnwidth]{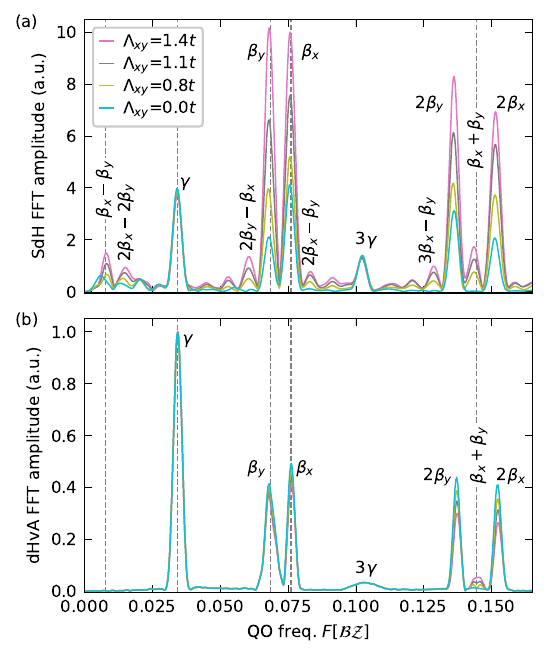}
    \caption{Numerically computed QO spectra for the parameters regime generating the FSs shown in Fig.~\ref{fig:2}~(a). In (a) we analyzed the conductance whereas in (b) we analyzed the density of states $\rho(\mu)$. The theoretical prediction for the three basis frequencies and the sum- and difference frequency, based on the area of the FSs, are indicated as grey dashed lines.}
    \label{fig:3}
\end{figure}


{\it Slow QO frequency from orbital selective scattering.--}
Finally, we evaluate the conductance in orbital magnetic fields through samples of sizes up to $400\times 400$ sites with orbital selective impurities within the nematic phase. The dominant SdH peaks in the Fourier spectrum, see Fig.~\ref{fig:3} panel (a), are set by the FSs $\beta_x,\beta_y,\gamma$ and higher harmonics thereof. The combination frequencies $\beta_x - \beta_y$ and $\beta_x + \beta_y$ are clearly visible and, additionally, a variety of subleading higher order terms appear whose strength depends on the strength of the impurity scattering. In the lower panel (b) we show the spectrum of the density of states, which corresponds to QO of thermodynamic observables like the de Haas-van Alphen (dHvA) effect. In contrast to the SdH effect, the slow difference frequency is absent in the dHvA effect. The reason is that the latter only depends on the scattering via the Dingle factor whereas scattering dominates transport \cite{Leeb_DiffFreq,Huber2023}, which is also confirmed by the strong (weak) dependence of the QO signals for the upper (lower) panels. Thus, a careful comparison between QO frequencies of SdH and dHvA can confirm our unusual QO without a FS orbit.

{\it Discussion and Conclusion.--} 
We have shown that a robust slow QO frequency emerges in minimal models of iron-chalcogenides. The key ingredients were the broken rotational symmetry between the electron pockets in the nematic phase and an efficient coupling between these pockets. The latter can originate from an orbital selective scattering, e.g. a dominating impurity contribution of the $d_{xy}$ orbital. We provided full numerical lattice calculations with orbital magnetic fields, which also confirm recent analytical works on difference frequency QOs without semiclassical orbits beyond the Onsager relation~\cite{Leeb_DiffFreq,Huber2023}. Further supporting evidence of our scenario is that the experimentally extracted masses from the temperature dependence of the QOs \cite{Coldea2019} is in accordance with our analytical predictions \cite{Leeb_DiffFreq}, namely the mass of the slow frequency roughly equals the difference of the ones of the electron pockets.

Of course, neither our effective two-band nor the three-orbital model (which is already challenging numerically) captures all details of the complicated electronic structure of iron-chalcogenides~\cite{coldea2018key}. In fact, we have neglected any correlation effects, which could further increase scattering between the electron pockets, e.g. by collective spin fluctuations. However, our scenario requires no preconditions except a finite coupling of the electron pockets via scattering. Therefore, we expect our scenario to be reproducible in any microscopic model of iron-chalcogenides. In summary, we argue that our results are a robust feature of the nematic phase of iron-chalcogenides and elucidate that no additional pocket of a nematic Lifshitz transition is required to explain the QO experiments~\cite{Coldea2019,Reiss2020}.

The correct assignment of QO frequencies with putative FS orbits is crucial for correctly identifying the electronic structure in iron-chalcogenides and beyond. Alas, our scenario of sharp QOs without FS orbits further complicates the interpretation of QO data. However, it also provides novel insights into subtle details of quasiparticle scattering  otherwise inaccessible  in experiments. 

We showed that the slow QO frequency of iron-chalcogenides can be explained by the presence of  orbital selective impurity scattering, which has implications for the SC pairing symmetry. It is normally expected that impurities, as necessarily present in heavily disordered FeSe$_{1-x}$S$_x$~\cite{teknowijoyo2016enhancement}, suppress s$^{\pm}$ superconductivity~\cite{efremov2011disorder,chubukov2012pairing}. However, the orbital selective scattering does {\it not } couple the electron and hole pockets, which would be detrimental for s$^{\pm}$ pairing. Thus, the new QO  mechanism possibly explains the robustness of superconductivity in the iron-chalcogenides. 

We hope that the observation and quantification of similar QO frequencies can lead to a more precise identification of the electronic structure of other correlated electron materials.

{\it Data and code availability.--}
Code and data related to this paper are available on Zenodo \cite{code} from the authors
upon reasonable request.

\begin{acknowledgments}
We acknowledge helpful discussion and related  collaborations with  N.~Huber, M.~Wilde and C.~Pfleiderer. We thank A. Chubukov, A. Coldea and T. Shibauchi, for helpful discussions and comments on the manuscript.
V.~L. acknowledges support from the Studienstiftung des deutschen Volkes. J.~K. acknowledges support from the Imperial-TUM flagship partnership. The research is part of the Munich Quantum Valley, which is supported by the Bavarian state government with funds from the Hightech Agenda Bayern Plus.
\end{acknowledgments}

\clearpage
\newpage
\appendix
\begin{widetext}

\section{3-orbital tight-binding model}
The Hamiltonian features nearest- and next-nearest-neighbor hoppings:
\begin{align}
	H_0 =& \sum_{\v{r}} \v{\Psi}^\dag_{\v{r}+\hat{\v{x}}}
	\begin{pmatrix}
		t_2 & 0 & t_7 \\
		0 & t_1 & 0 \\
		-t_7 & 0 & t_5
	\end{pmatrix} \v{\Psi}_{\v{r}}
	+
	\v{\Psi}^\dag_{\v{r}+\hat{\v{y}}}
	\begin{pmatrix}
		t_1 & 0 & 0 \\
		0 & t_2 & t_7 \\
		0 & -t_7 & t_5
	\end{pmatrix} \v{\Psi}_{\v{r}}
	+
	\v{\Psi}^\dag_{\v{r}+\hat{\v{x}}+\hat{\v{y}}}
	\begin{pmatrix}
		t_3 & -t_4 & t_8 \\
		-t_4 & t_3 & t_8 \\
		-t_8 & -t_8 & t_6
	\end{pmatrix} \v{\Psi}_{\v{r}}
	\nonumber\\&+
	\v{\Psi}^\dag_{\v{r}+\hat{\v{x}}-\hat{\v{y}}}
	\begin{pmatrix}
		t_3 & t_4 & t_8 \\
		t_4 & t_3 & -t_8 \\
		-t_8 & t_8 & t_6
	\end{pmatrix} \v{\Psi}_{\v{r}}
	+ \text{h.c.}
	+
	\v{\Psi}^\dag_{\v{r}}
	\begin{pmatrix}
		-\mu-\delta\mu & -\ii h_y & 0 \\
		\ii h_y & -\mu+\delta\mu & 0 \\
		0 & 0 & \Delta_{xy}-\mu
	\end{pmatrix} \v{\Psi}_{\v{r}}
\end{align}
Defining $\v{\Psi}_{\v{r}} = \frac{1}{\sqrt{N}} \sum_{\v{k}} \e^{-\ii \v{k} \v{r}}\v{\Psi}_{\v{k}}$ we obtain 
\begin{align}
	H_0 &= \sum_{\v{k}} \v{\Psi}_{\v{k}}^\dagger (T(\v{k})-\mu) \v{\Psi}_{\v{k}} + \delta \mu
	\begin{pmatrix}
		d_{\v{k}, xz} \\ d_{\v{k}, yz}
	\end{pmatrix}^\dag
	\sigma^z
	\begin{pmatrix}
		d_{\v{k}, xz} \\ d_{\v{k}, yz}
	\end{pmatrix}
\end{align}
where
\begin{align}
	T_{11}(k) &= 2t_2 \cos k_x + 2t_1 \cos k_y + 4t_3 \cos k_x \cos k_y \\
	T_{22}(k) &= 2t_1 \cos k_x + 2t_2 \cos k_y + 4t_3 \cos k_x \cos k_y \\
	T_{33}(k) &= 2t_5 (\cos k_x + \cos k_y) + 4t_6 \cos k_x \cos k_y+\Delta_{xy}\\
	T_{12}(k) &= T_{21}(k)^* = 4t_4 \sin k_x \sin k_y + \ii h_y\\
	T_{13}(k) &= T_{31}(k)^* = 2 \mathrm{i} t_7 \sin k_x + 4 \mathrm{i} t_8 \sin k_x \cos k_y \\
	T_{23}(k) &= T_{32}(k)^* = 2 \mathrm{i} t_7 \sin k_y + 4 \mathrm{i} t_8 \cos k_x \sin k_y. 
\end{align}
The values of the hopping parameters are shown in Tab.~\ref{tab:param}.

\begin{table}
	\centering
	\begin{tabular}{c|ccccccccccc}
		$\mac{M}$ & $t_1$ & $t_2$ & $t_3$ & $t_4$ & $t_5$ & $t_6$ & $t_7$ & $t_8$ & $\Delta_{xy}$ & $\mu$ & $h_y$ \\ \hline
		$\mac{C}$ & $0.2 t$ & $0.6 t$ & $0.3 t$ & $-0.1 t$ & $2 t$ & $3 t$ & $-2 t$ & $t$ & $4 t$ & $2.7 t$ & $0$
	\end{tabular}
	\caption{The tight-binding parameters used throughout this manuscript.}
	\label{tab:param}
\end{table}

\section{Numerical implementation and QO analysis}
We have implemented the tight-binding models and calculated the conductance and the density of states for finite magnetic fields using the python package kwant \cite{kwant}. The main methodological steps are shown in Fig.~\ref{fig:A1}. 

We compute the conductance $G$ with a 2-point measurement, see Fig.~\ref{fig:A1}~(a) via the built-in Landau--B\"uttiker algorithm, which is based on an S-matrix approach. For the 2-orbital model, we used a system size of $300\times 300$ lattice sites and for the 3-orbital model a system size of $400\times 400$ lattice sites.

In this work, we fit the $N_\text{data}(I_\Phi)$ data points inside an interval $2 \pi/\Phi \in I_\Phi$ with a 4th order polynomial. After subtracting the polynomial we scale the signal with a window and pad it symmetrically with $4 N_\text{data}(I_\Phi)$ zeros. Then the signal is Fourier transformed and we always show the absolute of the Fourier transformed signal. For Fig.~\ref{fig:1}~(c) and (d) we used $I_\Phi = [20,250]$ and a Hamming window. For Fig.~\ref{fig:1}~(e) we used $I_\Phi = [40,300]$ and a Hamming window. For Fig.~\ref{fig:1}~(e) we used $I_\Phi = [40,300]$ and a Hamming window. For Fig.~\ref{fig:3}~(a) we used $I_\Phi = [30,720]$ and a Blackman--Harris window. 

We compute the density of states $\rho$ with the kernel polynomial method \cite{KPM}. For sampling the spectral density, we use 30 randomly chosen vectors such that only the bulk of the system is sampled. Cutting off the edges suppresses edge state effects over bulk effects, simulating the thermodynamic limit. We define the bulk by the set of all lattice points at least 50 sites away from the edges. We used 7000 Chebyshev moments. In Fig.~\ref{fig:3}~(b) we analyzed $\rho(\omega=0,\Phi)$ inside the interval $I_\Phi = [90,720]$ using a Hamming window.

Note for experimental comparison that $B = \frac{\Phi}{2\pi} \frac{\hbar}{e a^2} = \frac{\Phi}{2\pi} \times 4.63$kT for a lattice constant $a=3.77$\r{A} of iron selenides~\cite{Margadonna2008_crystal}. We work in units where $e=\hbar=a=1$. Hence, the analyzed $I_\Phi$ intervals translate to roughly $6$ - $46$T.

\begin{figure}
	\centering
	\includegraphics[width=\textwidth]{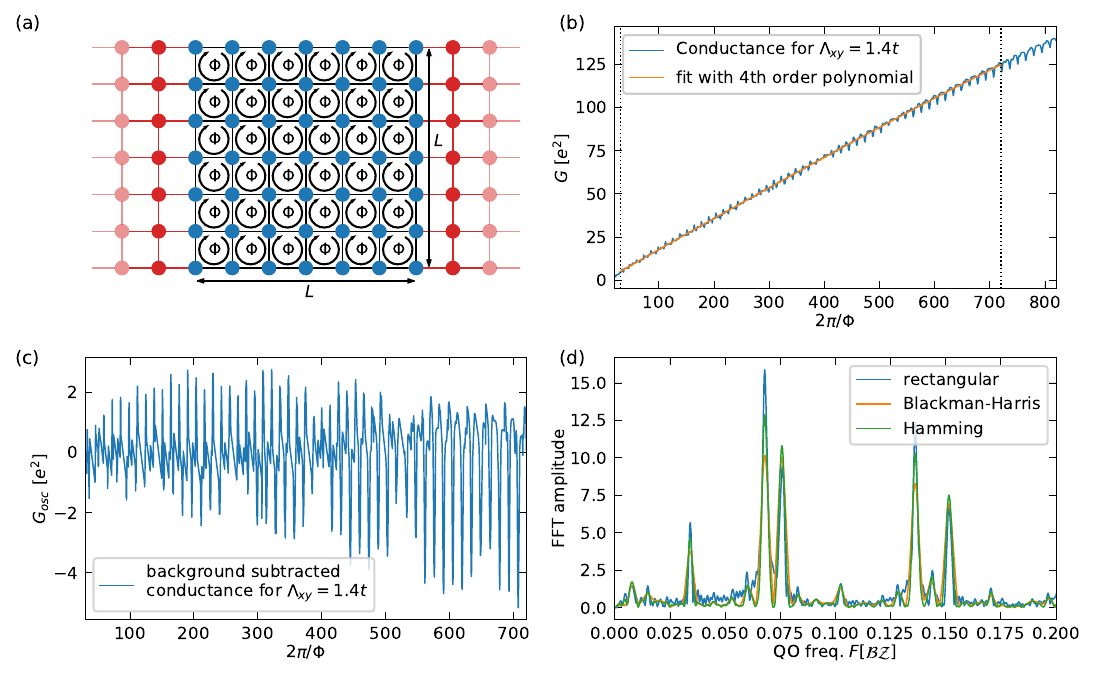}
	\caption{Panel (a) shows a schematic sketch of the implemented model for conductance simulations. The scattering region (blue vertices) of system size $L\times L$ implements the tight binding model with impurities and each hopping is scaled with a Peierls phase such that flux per plaquette is $\Phi$. The leads (red vertices) are translational invariant, hence are not exposed to the magnetic field and do not have impurities. The conductance $G$ from the left lead to the right lead is computed. Panel~(b) shows the exemplary result for the conductance $G$ of the 3-orbital model, i.e. the data shown in Fig.~\ref{fig:3}. We fit the conductance in the field region $I_\Phi$ (marked by the black dotted lines) with a $4^\text{th}$-order polynomial which is then subtracted from the numerical data to obtain the oscillating part of the signal, see panel~(c). To obtain a spectrum, panel~(d), the oscillating signal is zero padded, scaled with a window and then Fourier transformed. The identified peaks are consistent for different types of chosen windows.}
	\label{fig:A1}
\end{figure}

\end{widetext}

\bibliography{bib}

\end{document}